\documentclass[fleqn,10pt]{wlscirep}
\usepackage[utf8]{inputenc}
\usepackage[T1]{fontenc}

\usepackage{graphicx}
\usepackage{hyperref}
\usepackage{tikz,subfigure}  
\usetikzlibrary{arrows,shapes,positioning,shadows,svg.path,backgrounds,fit}

\newcommand{\op}[1]{\hat{#1}}

\newcommand{\qmean}[1]{\langle{#1}\rangle}

\title{Heat capacities of thermally manipulated mechanical oscillator at strong coupling}

\author[1,*]{Michal Kol\'a\v r}
\author[2]{Artem Ryabov}
\author[1]{Radim Filip}
\affil[1]{Palack\'{y} University, Department of Optics, 17.~listopadu 1192/12, 771~46~Olomouc, Czech Republic}
\affil[2]{Charles University, Faculty of Mathematics and Physics, Department of Macromolecular Physics, V~Hole{\v s}ovi{\v c}k{\' a}ch~2, 180~00 Praha, Czech Republic}

\affil[*]{kolar@optics.upol.cz}

\begin{abstract}
Coherent quantum oscillators are basic physical systems both in quantum statistical physics and quantum thermodynamics. Their realizations in lab often involve solid-state devices sensitive to changes in ambient temperature. We 
represent states of the solid-state optomechanical oscillator with temperature-dependent frequency by equivalent states of the mechanical oscillator with temperature-dependent energy levels. We interpret the temperature dependence as a consequence of strong coupling between the oscillator and the heat bath. We explore parameter regimes corresponding to anomalous behavior of mechanical and thermodynamic characteristics as a consequence of the strong coupling: (i) 
The localization and the purification induced by heating, and (ii) the negativity of two generalized heat capacities. The capacities can be used to witness non-linearity in the temperature dependency of the energy levels. Our phenomenological experimentally-oriented approach can stimulate development of new optomechanical and thermomechanical experiments exploring basic concepts of strong coupling thermodynamics.
\end{abstract}
\begin{document}
\flushbottom
\maketitle
%
\thispagestyle{empty}

\section{Introduction}

Quantum optomechanics and electromechanics become physical bridges between developed atomic, molecular and optical physics and emerging quantum thermodynamics. It is due to possibility to operate mechanical and electrical oscillators as quantum systems\cite{aspelmeyer2014} as well as to understand them as a part of thermodynamic processes and engines.\cite{zhangPRL2014} The solid-state nature of such oscillators simultaneously implies that their basic characteristics (such as mode frequencies) can be strongly influenced by ambient temperature.\cite{Zadeh2010,Zadeh2006, Zaitsev2012, Yuvaraj2013, Khanaliloo2015, Jungwirth2016} Despite the fact that this sensitive dependence can make them promising temperature sensors,\cite{Pandey2010} the oscillators can represent first experimental demonstrations of non-linear temperature-driven dynamics.\cite{SilerSciRep2017, SilerPRL2018} On a more fundamental level, the controllable temperature-dependence of oscillator parameters opens doors towards experimental investigation of thermodynamics of quantum mechanical systems strongly coupled to their environment. 

At the same time, thermodynamics of systems at strong coupling has recently attracted a significant attention and a number of general theoretical studies has appeared,\cite{deMiguel2015, Jarzynski2004, GelinPRE2009, EspositoNJP2010, PhilbinNJP2016, JarzynskiPRX2017, StrasbergPRE2017, SeifertPRL2016, TalknerHanggiPRE2016} which generalized classical theoretical works,\cite{rushbrooke,  Radkowsky1948, Fan1951, Elcock1957, EminPRB1984, ALDEA1989375, Donnelly1977} and several  particular models with temperature-dependent energy levels have been explored.~\cite{ShentalEPL2009,  CARDONA2014680, Takuya2016, Villegas2016,deMiguel2017JPCB}  However, the fundamental theoretical approaches encounter severe difficulties already at the level of definitions of basic thermodynamic quantities,\cite{TalknerHanggiPRE2016} where the concepts of heat and entropy production cannot be unambiguously identified. A promising way out of these theoretical struggles is, in our opinion, to propose and  build in lab real-world quantum-mechanical thermodynamic devices. Such experiments would inspire an operational theoretical approach to thermodynamics at strong coupling. The operationally defined quantities created to describe particular new effects in individual mechanical systems will not suffer from ambiguities present in general analysis and, in turn, they may help to establish a new general paradigm in the field.

In the present work we make first steps in development of the experimentally-oriented approach to stochastic thermodynamics at strong coupling inspired by the optomechanical experiments. We start from a typical outcome of such experiments where measured quantities are spectra and the phonon number distribution of the optomechanical mode with the frequency $\omega=\omega(T)$. The temperature dependence of the mode frequency,  natural in solid-state oscillators,\cite{Zadeh2010,Zadeh2006, Zaitsev2012, Yuvaraj2013, Khanaliloo2015, Jungwirth2016}  allows us to explore effects of strong coupling, which are not present in weakly coupled oscillators with a constant frequency. In particular, we discuss the temperature-induced localization (Sec.~\ref{sec:mechanics}), purification (Sec.~\ref{sec-entropy}), and propose two experiments described by corresponding generalized heat capacities (Sec.~\ref{sec-capacity}): The first utilizes a reconstruction procedure of the system density matrix,\cite{aspelmeyer2011PNAS} whereas the second one  is an analogue of the differential scanning calorimetry.\cite{Bower}

\begin{figure}[t!]
\centering 
\includegraphics[width=.95\linewidth]{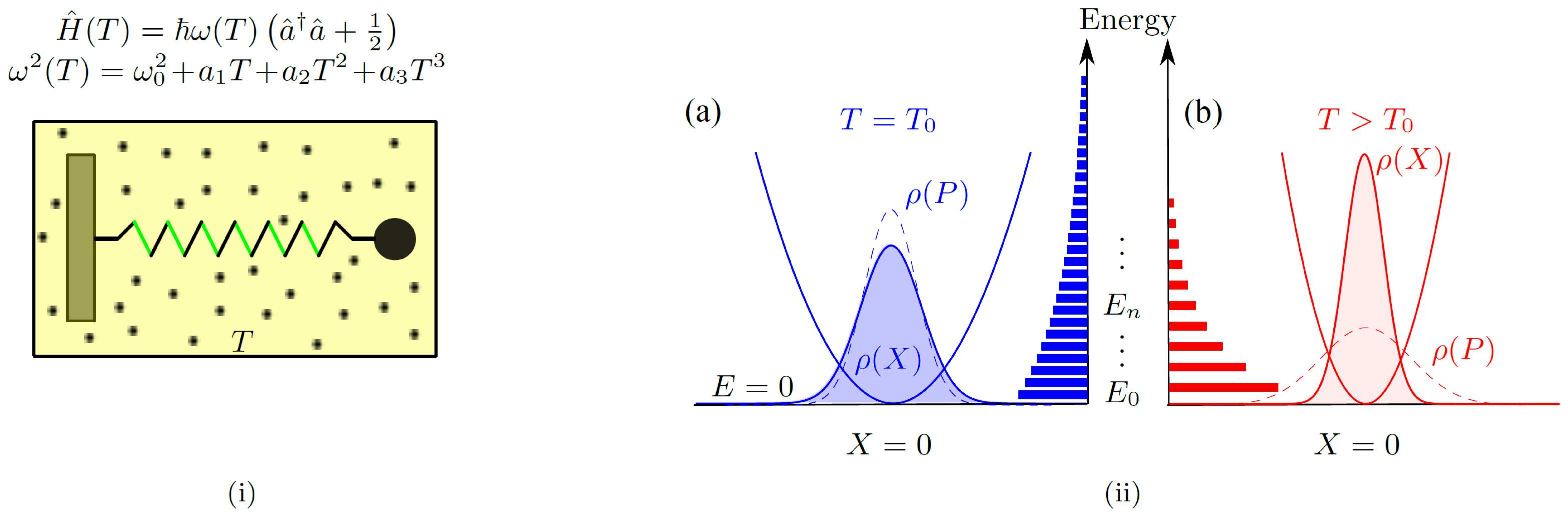}
\caption{
(i) Mode frequencies of solid-state optomechanical oscillators are sensitive to changes in ambient temperature. This temperature-dependence can be utilized to model effects of strong coupling. Using the position and momentum $\op{X}(T)$ and $\op{P}(T)$ defined in Eqs.~\eqref{eq-oscill-X} and~\eqref{eq-oscill-P}, respectively, outcomes of optomechanical experiments can be naturally interpreted and physically understood based on the representative mechanical toy model: the harmonic oscillator with a structured spring (alternating green/black segments), which stiffness depends on the heat bath temperature $T$. 
 Remarkable physical effects induced by the dependence $\omega=\omega(T)$ are main topic of the present work. The standard harmonic oscillator characterized by a constant frequency is not able to demonstrate all effects described in the main text and depicted on panel (ii). (ii) Thermal change of the oscillator potential: (a) The oscillator in contact with a heat bath
at lower temperature $T_0$ (blue), initially. (b) The temperature of the heat bath is increased to $T >T_0$, (red). The spectrum and the oscillator state change, showing the reduction of the position variance ({\it localization} of the position distribution $\rho(X)$, shaded) with increasing temperature, and the spreading of the momentum distribution $\rho(P)$ (dashed), discussed in Sec.~\ref{sec:mechanics}. Such behavior should not be misinterpreted as squeezing of the oscillator thermal state.\cite{RashidPRL117}
The horizontal bars represent populations of energy eigenstates showing the reduction of the von Neumann entropy ({\it purification} of the state) with increasing temperature, cf.\ Sec.~\ref{sec-entropy}. These effects are responsible for the negativity of capacities discussed in Sec.~\ref{sec-capacity}.} \label{fig-model}
\end{figure}

From the general perspective, the approach adopted here goes along the lines known in the stochastic thermodynamics\cite{SeifertRPP2012} where perfectly controllable experiments with optical tweezers are used to simulate and verify basic theoretical predictions.\cite{CilibertoPRX2017} Here, instead of the optical tweezers, the single mode of the solid-state optomechanical oscillator is used as an experimental simulator for the mechanical harmonic oscillator with temperature-dependent frequency  $\omega(T)$, see Fig.~\ref{fig-model}. There are two main reasons why we focus on optomechanical platforms in our considerations. Firstly, they exhibit the aforementioned temperature dependence of $\omega(T)$, which can be understood as a consequence of the strong coupling of the mechanical oscillator (with the bare frequency $\omega_0$) to the thermal environment, cf.\ the definition~\eqref{eq-omegaT}. Secondly, the light defining the optical part of the system can be conveniently used to measure all the important characteristics of the mechanical part described in this paper.\cite{HoferPRA2011,VannerNCom2013,WieczorekPRL2015} This is the sole purpose of the light in our considerations and we do not assume it to affect the discussed mechanical and thermodynamic properties in any sense in the following.

For energies of the optomechanical mode we assume
\begin{eqnarray}
E_{n}(T)=\hbar\omega(T) \left( n+\frac{1}{	2}\right), \qquad n=0,1,2,\ldots ,
\label{eq-total-Hamiltonian-diag}
\end{eqnarray}
with the mode frequency written as  
\begin{equation}
\label{eq-omegaT}
\omega^{2}(T)=\omega^{2}_0+f(T).
\end{equation}
Above, $\omega_0$ stands for the bare oscillator frequency (a temperature-independent constant) and the function $f(T)$ represents a frequency shift because of the strong coupling to the environmental degrees of freedom. We assume the polynomial form of this function 
\begin{equation} 
\label{eq-omega}
f(T)= \sum_{k=1}^\Omega a_kT^k,
\end{equation}
with $\Omega$ being the maximum order up to which we truncate the series expanding $f(T)$ in $T$. In modeling optomechanical oscillators, frequently only weak linear dependence on temperature is considered,\cite{Zaitsev2012} which is valid for small frequency shifts. Yet, for large frequency shifts the dependence of frequency on temperature is usually non-linear.\cite{Pandey2010, Mitropolskii, Giessibl1997, Holscher2001, Grabert1984} 
In the present work, we drop the linearization assumption and focus on effects induced by higher order terms in Eq.~\eqref{eq-omega}. In discussed examples we assume the maximum order of the temperature dependence in the range $0<\Omega \leq 3$. We show that such non-linear temperature dependence can cause reduction of the position variance, and/or of the von Neumann entropy with increasing $T$, and the negativity of corresponding heat capacities.

The temperature-dependent spectrum~\eqref{eq-total-Hamiltonian-diag} can be interpreted as eigenvalues of the so-called Hamiltonian of the mean force (HMF), $\op{H}(T)$, introduced in Secs.~\ref{sec:model} and \ref{sec:phenomenological}. Sec.~\ref{sec:mechanics} discusses the mechanical equilibrium properties of the system, namely the temperature dependence of the position variance of the oscillator. In Sec.~\ref{sec-entropy}, we study the temperature dependence of the von Neumann entropy, focusing on its decrease with increasing bath temperature. Section~\ref{sec-capacity} analyses the temperature dependence of two thermodynamic coefficients and their mutual connection. In Sec.~\ref{sec:conclusions}, we summarize experimental conditions necessary to observe the predicted results.

\section{Equivalent mechanical oscillator}
\label{sec:model}

The spectrum~\eqref{eq-total-Hamiltonian-diag} represents eigenvalues of the operator 
\begin{equation} 
\op{H}(T) = \hbar\omega(T)\left(\op{a}^\dagger\op{a}+\frac{1}{2}\right),
\label{eq-oscill-ham}
\end{equation} 
with $ \op{a}^{\dagger}$ and $\op{a}$ being respectively the creation and the annihilation) operators of the mode $\omega(T)$. In optomechanical experiments, the mechanical mode statistics is reconstructed from the homodyne detection signal measured on the light interacting with the mechanical mode. The result of such measurements can be used to characterize the state of a representative  mechanical oscillator in terms of linear combinations of $\op{a}^{\dagger}$ and $\op{a}$ called quadratures. \cite{aspelmeyer2014,WieczorekPRL2015} Two quadratures particularly suitable for our purposes read 
\begin{eqnarray}
\label{eq-oscill-X}
\op{X}(T)&=& \sqrt{\frac{\hbar}{2\omega(T)}}\left(\op{a}^\dagger+\op{a}\right),\\
\op{P}(T) &=& i\sqrt{\frac{\hbar\omega(T)}{2}}\left(\op{a}^\dagger-\op{a}\right).
\label{eq-oscill-P}
\end{eqnarray} 
The quadratures $\op{X}(T)$ and $\op{P}(T)$ represent respectively the position and the momentum operators of the quantum linear harmonic oscillator with the mass $m=1$ and frequency $\omega(T)$. The Hamiltonian of such oscillator follows directly from Eq.~\eqref{eq-oscill-ham}. It reads 
\begin{equation} 
\op{H}(T) = \frac{\op{P}^2}{2}+\frac{1}{2} \left[\omega_0^2+f(T) \right] \op{X}^2. 
\label{eq-oscill-ham2}
\end{equation} 
To test the observability of considered phenomena, a simulation where both temperature and frequency are tuned in a correlated way can be the first step. Any optomechanical platform can, after proper interpretation, be used to simulate the mechanical oscillator just by choosing the proper quadratures. 

Some experimental platforms \cite{aspelmeyer2014} and theoretical models \cite{Grabert1984} can also exhibit an effective change of the mass with temperature. In this case, another possibility is to define the mechanical quadratures as 
$ \op{x}(T)=\left(\op{a}^\dagger+\op{a}\right)\sqrt{\hbar/2m_{\rm eff}(T)\omega(T)} $, and 
$\op{p}(T)=i\left(\op{a}^\dagger-\op{a}\right) \sqrt{\hbar m_{\rm eff}(T)\omega(T)/2} $, 
with $m_{\rm eff}(T)$ being the temperature-dependent effective oscillator mass. 
In  practice, the effective mass is always determined indirectly from finite element modeling of the optomechanical platform and using the equality of energy of the cavity elastic vibrations and energy of a 1D harmonic motion.\cite{Pinard1999,Zadeh2006,aspelmeyer2014} However, we would like to point out that this cumbersome indirect identification of $m_{\rm eff}(T)$ is not necessary, because all equilibrium thermodynamic characteristics do not depend on $m_{\rm eff}(T)$ but only on $\omega(T)$. Therefore, the simple definitions of $\op{X}(T)$ and $\op{P}(T)$ in Eqs.~\eqref{eq-oscill-X} and \eqref{eq-oscill-P}, respectively, are sufficient to fully explore and interpret all important effects of strong coupling in equilibrium. Therefore, we can rescale the position and momentum in such way, that consequent analysis is the same as for the case of thermally independent mass.

\section{Simulation of strong coupling thermodynamics}
\label{sec:phenomenological}
The total Hamiltonian of the bare oscillator interacting with
the heat bath reads
\begin{equation} 
\op{H}_{\rm tot} = \op{H}_0+\lambda\op{H}_{\rm I}+\op{H}_{\rm B},
\label{eq-total-ham}
\end{equation} 
where the dimensionless parameter $\lambda$ reflects the system-bath coupling strength. When the interaction $\lambda\op{H}_{\rm I}$  between the bare oscillator and the equilibrium heat bath is weak, the oscillator
is known to thermalize into the Gibbs canonical state, $\op{\rho}_{\rm G}(T)\sim \exp\left[-\op{H}_0/k_{\rm B}T \right]$. On the other hand, for a strong coupling, the state of the oscillator will depend on parameters of the ambient environment, even in the long-time limit. However, even for the strong coupling case, the equilibrium state of the oscillator can still be formally considered in the canonical form determined by the bath temperature $T$,
\begin{equation}
\label{eq-Gibbs}
\op{\rho}(T) = \frac{1}{Z(T)}\exp\left[-\op{H}(T)/k_BT\right],
\end{equation}
where the operator $\op{H}(T)$ is known as the Hamiltonian of mean force (HMF), \cite{TalknerHanggiPRE2016,SeifertPRL2016,JarzynskiPRX2017} and $Z(T)$ is the partition function, $Z(T)={\rm Tr}\left[\exp\left[-\op{H}(T)/k_BT\right]\right]$. The term HMF is derived from the potential of mean force originally introduced in theory of fluids.\cite{Kirkwood1935, ROUX19991}

The HMF is obtained after averaging over bath degrees of freedom in the total system as follows \cite{TalknerHanggiPRE2016,SeifertPRL2016,JarzynskiPRX2017}
\begin{equation}
{\rm e}^{-\op{H}(T)/k_BT}= \left< {\rm e}^{-\op{H}_{\rm tot}/k_BT} \right>_{\rm B}.
\label{eq-exp-op-def}
\end{equation}
Equation~\eqref{eq-exp-op-def} represents a general definition of the exponential operator describing the subsystem of interest. It is valid for quantum, as well as for classical systems. Following this definition, application of a logarithm on both sides of \eqref{eq-exp-op-def} leads to the result for the quantum HMF
\begin{equation}
\op{H}(T)\equiv -k_BT\ln\left< {\rm e}^{-\op{H}_{\rm tot}/k_BT} \right>_{\rm B}.
\label{eq-HMF-op-def}
\end{equation} 
The structure of this operator can be quite complex in the general quantum case, due to the non-commutativity of respective terms in $ \op{H}_{\rm tot}$, Eq.~\eqref{eq-total-ham}. Significant simplification is possible in situations in which classical description of the system is possible, i.e. all terms in $ \op{H}_{\rm tot}$, {\it commute} and we can split the exponential operator into classical parts containing solely $H_0$, $H_{\rm B}$, and $H_{\rm I}$. This is the reason why, in the classical (high phonon number) regime, where most current optomechanical devices operate,\cite{aspelmeyer2014} we can split the HMF in two terms
\begin{equation}
\label{eq-Hmf}
H(T) = H_0 - k_BT \ln \left< {\rm e}^{-\lambda H_{\rm I}/k_BT} \right>_{\rm B}.
\end{equation}
The second term on the right-hand side of Eq.~\eqref{eq-Hmf} contains all details of the interaction between the system and bath and depends on the parameters of the bath itself (e.g., spectrum of its phonon modes). 
Explicit first-principle calculations of thermodynamic functions for a strongly coupled oscillator can be found in the literature,\cite{FordPRL1985, FORD1988270, FordPRB2007} as well as discussion of specific heats. \cite{HanggiIngoldTalknerNJP2008, IngoldHanggiTalknerPRE2009}

Comparing Eq.~\eqref{eq-Hmf} with the Hamiltonian~\eqref{eq-oscill-ham2} we can identify the latter with the HMF of the real mechanical oscillator strongly coupled to its environment. The temperature dependent shift of the mode frequency $\omega(T)$, represented by the function $f(T)$ in the definition~\eqref{eq-omega}, can be understood as a consequence of the strong coupling. In this way the optomechanical experimental platform can be used to simulate and explore physics of the actual oscillator with the HMF~\eqref{eq-Hmf}.


\section{Temperature-induced localization}
\label{sec:mechanics}
Thorough understanding of mechanical characteristics is crucial for a physical insight into different equilibrium thermodynamic properties of states \eqref{eq-Gibbs}, discussed in the next section, modified by temperature-dependent frequency shift as given in Eq.~\eqref{eq-oscill-ham}. In this section we discuss the behavior of such characteristic, namely  the variance of the quadrature $\op{X}$, ${\rm Var}_X(T)$. For different forms of the frequency shift function $f(T)$, defined in Eq.~\eqref{eq-omega}, we obtain qualitatively different temperature dependencies of the position variance. For certain parameters, we observe the position {\it localization} when increasing $T$. In Sec.~\ref{sec-entropy}, the localization will be generalized to the {\it purification} of the state, i.e. the entropy reduction, where its measurement and the thermodynamic consequences will be discussed.  Interestingly, the localization does not necessary implies the purification of the state.

For the thermal state~\eqref{eq-Gibbs}, the quantum mechanical result for the variance of the quadrature $\op{X}$ reads
\begin{equation}
{\rm Var}_X(T)=\frac{\hbar}{\omega(T)}\left( \qmean{\op{n}}+ \frac{1}{2} \right),
\label{eq-mech-variance-general}
\end{equation}
where $\qmean{\op{n}}=[\exp(\hbar\omega(T)/k_BT)-1]^{-1}$ is the Bose-Einstein thermal population of the oscillator mode. Therefore, the temperature-dependent  frequency enters both the thermal population and the ground state variance for $\langle\hat{n}\rangle =0$. It makes a localization of the system nonlinear even for linear change of $\omega(T)$ over the temperature.

In the current experiments, temperature of environment keeps the oscillator in the equilibrium still far away from the ground state.\cite{aspelmeyer2014}  Therefore we primarily focus on a high-temperature limit $\hbar\omega(T)/k_BT\ll 1$. In this limit, keeping $m=1$ as explained in Sec.~\ref{sec:model}, variances of position and momentum are given by 
\begin{equation}
{\rm Var}_X(T)\approx  \frac{k_BT}{\omega^2(T)}, \quad 
{\rm Var}_P(T)\approx  k_BT,
\label{eq-mech-variance-classical}
\end{equation}  
respectively. Apparently, the momentum variance ${\rm Var}_P(T)$ does not depend on $\omega(T)$ and it behaves just as for the system weakly coupled to a bath. It however depends on temperature $T$ and therefore, it will influence other quantities, e.g. entropy, when $T$ changes. If the frequency  does not depend on temperature, $\omega(T)=\omega_0$, the variance ${\rm Var}_X(T)$ increases linearly with $T$, and the oscillator exhibits {\em thermal delocalization}. 

Eqs.~\eqref{eq-mech-variance-classical} are valid in the high-temperature limit $\hbar\omega(T)/k_BT\ll 1$. Our system reaches this limit and stays there for appropriate parameter values in equation~\eqref{eq-omega}: $\Omega\leq 3$, $a_\Omega <0$, or  $\Omega\leq 2$, $a_\Omega >0$ for almost any $T$, except the region of extremely small bath  temperatures (not attainable in current optomechanical and electromechanical experiments \cite{aspelmeyer2014}). This behavior is clear for $a_\Omega\leq 0$. For $a_\Omega > 0$ it is caused by the decrease of the ratio $\omega(T)/T$ for increasing temperature of the bath. Even-though the energy level separation $\hbar\omega(T)$ increases with $T$, the average thermal energy $k_BT$ increases faster, thus still allowing for the average thermal population increase.
On contrary, in the case $a_3>0$ the relative increase of $\omega(T)/T$ might bring the oscillator out of Eq.~\eqref{eq-mech-variance-classical} validity regime for increasing temperature. Physically, this corresponds to the situation in which $\omega(T)$ increases fast enough, so that the bath is less able to excite the temperature dependent inter-level energy difference $\hbar\omega(T)$ with the typical thermal energy $k_BT$. Such situation may bring the oscillator effectively closer to the ground state for increasing temperatures. This case, however, does not appear for the experimentally motivated parameters used in the present work.

\begin{figure}[t!]
\centering 
\includegraphics[width=.55\linewidth]{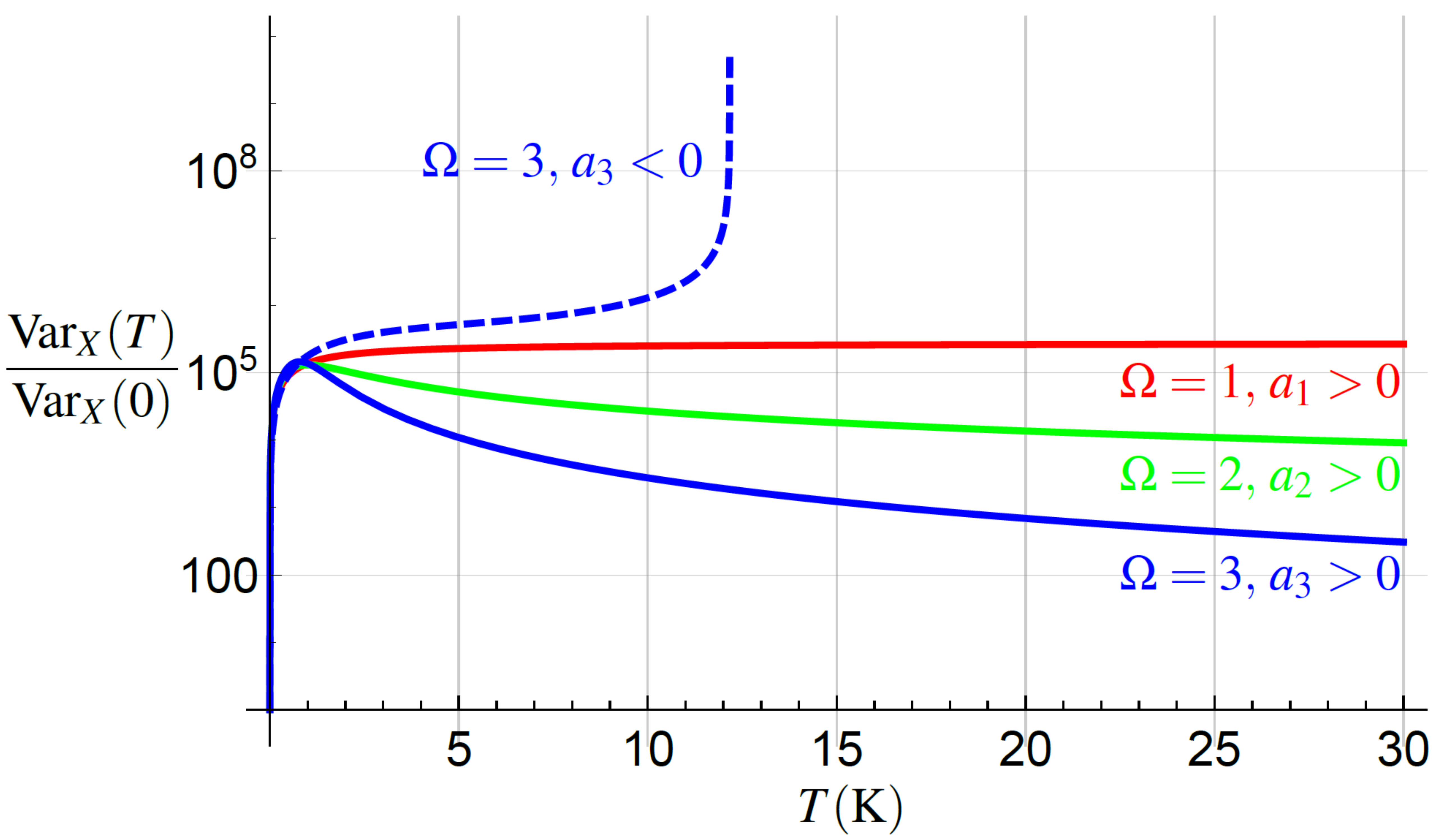}
\caption{The semi-logarithmic plot of the normalized variance of the oscillator position, Var$_X(T)$/Var$_X(0)$ Eq.~\eqref{eq-mech-variance-general}, for different values of the parameters in Eq.~\eqref{eq-omega}. For all curves the  parameters plotted correspond to the deep classical regime, $\hbar\omega(T)/ k_BT\ll 1$ (parameters inspired by recent experiments~\cite{aspelmeyer2011}). For negative coefficients $a_\Omega<0$ the variances diverge for large enough $T$, as illustrated by the blue-dashed curve for typical example $a_3<0$. The behavior for $a_\Omega >0$ is shown by full lines. The red and green curves correspond asymptotically (for large enough $T$) to the deep classical regime of the oscillator, $\hbar\omega(T)/k_BT\ll 1$, while having different asymptotics. The red curve saturates, while the green decreases as $\sim T^{-1}$. It witnesses {\em thermal steady-state localization}.
The blue-full curve corresponds to $\Omega=3,\;a_3>0$. 
In the regime plotted, the variance decreases as ${\rm Var}_X(T)\sim T^{-2}$. The parameters values used in these plots are $\omega_0\approx 10^6$ rad/s, $|a_\Omega T^\Omega|\approx \omega_0^2$ for all cases.} \label{fig-mech-all}
\end{figure}

For  $\Omega\leq 3$, $a_\Omega<0$, the variance \eqref{eq-mech-variance-classical} diverges at $T_{\rm d}\approx \sqrt[\Omega]{\omega_0^2/|a_\Omega|}$, see the dashed line in Fig.~\ref{fig-mech-all}. Such behavior is caused by of $\omega(T_{\rm d})=0$, meaning the confining potential disappears at $T=T_{\rm d}$, while the oscillator becomes a free particle.
Qualitatively different effects can be found for $\Omega\leq 2,\;a_\Omega>0$. 

For $\Omega=1,\;0<a_1$, the variance saturates at ${\rm Var}_X(T)=k_B/(a_1)$ for large enough temperatures, see the red curve in Fig.~\ref{fig-mech-all}. In this case, the temperature dependence of $\omega^2(T)$ becomes linear for high enough temperatures satisfying $a_1T\gg\omega_0^2$, canceling exactly the numerator in ${\rm Var}_X(T)$. ${\rm Var}_X(T)$ never exceeds this limiting value.

In the case $\Omega=2,\;0<a_2\ll k_B^2/\hbar^2$ we find ${\rm Var}_X(T)$ decreasing as ${\rm Var}_X(T)\sim T^{-1}$ for large enough temperatures satisfying $a_2T^2\gg\omega_0^2$, see the green line in Fig.~\ref{fig-mech-all}. This is caused by the quadratic increase of the denominator of \eqref{eq-mech-variance-classical} and linear increase of its nominator. Interestingly, this behavior does not oppose the fact that the oscillator still remains in the classical domain $\hbar\omega(T)/k_BT\ll 1$ due to $\omega(T)/T\sim 1$ for $a_2T^2\gg\omega_0^2$. Despite the oscillator stays in the classical regime, we observe an anomalous behavior -- {\em thermal localization} due to strong coupling with the heat bath.

The value $\Omega=3$ reveals yet another qualitatively different behavior. We find again ${\rm Var}_X(T)\rightarrow 0$, being approached as ${\rm Var}_X(T)\sim T^{-2}$, see the blue line in Fig.~\ref{fig-mech-all}, due to the cubic increase of the denominator of \eqref{eq-mech-variance-classical} and linear increase of its nominator. On contrary to the previous case $\Omega=2$, the oscillator aims to the quantum regime $\hbar\omega(T)/k_BT\approx 1$ due to the dependence $\omega(T)/T\sim T^{1/2}$ for $a_3T^3\gg\omega_0^2$. 

By inspection, Var$_X(T)$ is an increasing function of $T$ for low temperatures, meaning
\begin{eqnarray}
\frac{{\partial\;{\rm Var}}_X(0) }{{\partial}T}> 0.
\label{eq-mech-var-increase}
\end{eqnarray}
Eq.~\eqref{eq-mech-var-increase} and the oscillator localization (in certain cases) for large enough temperatures predicts the existence of a local maximum for the oscillator variance. By definition the maximum variance can not be overcome, in contrast to the common thermal delocalization. The derivative of the position variance in the classical limit \eqref{eq-mech-variance-classical} with respect to temperature $T$ reads
\begin{eqnarray}
\frac{{\partial\;{\rm Var}}_X(T) }{{\partial}T}=\frac{k_B(\omega_0^2-a_2T^2-2a_3T^3)}{\omega^4(T)},
\label{eq-mech-var-derivative}
\end{eqnarray}
where we have used the notation from Eq.~\eqref{eq-omega}. Looking for existence of the Var$_X(T)$ extreme amounts to finding the roots of the polynomial in the nominator of Eq.~\eqref{eq-mech-var-derivative}. 

We note that for $\Omega=1$ the coefficient $a_1$ does not appear in the nominator of Eq.~\eqref{eq-mech-var-derivative}, thus it can only saturate for $a_1>0$ as $T$ increases and can not affect the existence of the extrema. This means the extreme does not exist in this case, see Fig.~\ref{fig-mech-all}. For $a_1<0$, Eq.~\eqref{eq-mech-var-derivative} is always positive and Var$_X(T)$ diverges for $T_{\rm d}= \sqrt{\omega_0^2/|a_1|}$, the temperature for which $\omega(T_{\rm d})=0$, i.e., the confining potential disappears.

For $\Omega=2$ the positive root $T_{\rm max}>0$ of Eq.~\eqref{eq-mech-var-derivative} nominator exists only for $a_2>0$ and reads $T_{\rm max}=\sqrt{\omega_0^2/a_2}$. The maximum value of Var$_X(T)$ is then 
\begin{eqnarray}
{\rm Var}_X(T_{\rm max})=\frac{k_B}{a_1+2\sqrt{a_2\omega_0^2}},\label{eq-mech-V2-max}
\end{eqnarray}
meaning that in this regime the position variance can not globally exceed the value of Eq.~\eqref{eq-mech-V2-max}. Notice, $T_{\rm max}$ increases linearly with frequency, therefore such maximum may be found at high temperature for current high-frequency oscillators.\cite{aspelmeyer2011, PhysTod2012} Simultaneously, with increasing frequency the maximum variance \eqref{eq-mech-V2-max} decreases, therefore, for its determination the ability to measure small fluctuations of the mechanical oscillator is  required. 

In the case $\Omega=3$ finding the roots of Eq.~\eqref{eq-mech-var-derivative} nominator is straightforward but cumbersome. We use here the result for a simplified situation $a_2\ll a_3T$, yielding positive root $T_{\rm max}\approx \sqrt[3]{\omega_0^2/2|a_3|}$. This result is approximately valid if the maximum is reached still in the classical region of parameters. For $a_3<0$ the maximum does not exist, Var$_X(T)$ increases monotonically, and diverges for such $T_{\rm d}$ for which $\omega(T_{\rm d})=0$, see the blue-dashed line in Fig.~\ref{fig-mech-all}. 

For the last two cases we encounter the aforementioned effect of the {\it thermal localization} of the oscillator, i.e. the reduction of its position variance, with increasing the bath temperature $T$. The results of this paragraph suggest that for $\Omega=3$, $a_3>0$, the maximum of the position variance scales with the system frequency as ${\rm Var}_X(T_{\rm max})\sim \omega_0^{-4/3}$. It is thus preferable to work with higher frequency oscillators if lower values of position variance are the figure of merit.

\section{Temperature-induced purification}
\label{sec-entropy}

In this section we focus on the parallels and remarkable differences between the behavior of the Var$_X(T)$, reflecting the position {\it localization}, and the reduction of von Neumann entropy, $S(T)$ (denoted simply as the entropy from now on), of the state \eqref{eq-Gibbs}. The entropy defined in the standard way reads
\begin{equation}
S(T)=-k_B{\rm Tr}[\op{\rho}(T)\ln\op{\rho}(T)].
\label{eq-therm-S}
\end{equation}
The value of $S(T)$ reflects the purity of the oscillator state.\cite{ZurekRevModPhys2003,ColesRevModPhys2017} Contrary to the position uncertainty quantified solely by ${\rm Var}_X(T)$, the state purity depends on both ${\rm Var}_X(T)$ and  ${\rm Var}_P(T)$. To observe the temperature-induced purification (the decrease of $S(T)$ with $T$), the decrease of uncertainty in the position has to be faster than the increase of momentum uncertainty in the product ${\rm Var}_X(T){\rm Var}_P (T)$, completely determining the entropy $S(T)$. To observe such an unusual effect, a more strict condition on $\omega(T)$ must be satisfied compared to the thermally induced localization. In experiment, it requires a precise estimation of the density matrix $\op{\rho}(T)$, which can be done with the help of the homodyne detection.\cite{browleyNCom2016} 

The approximate result obtained from Eq.~\eqref{eq-therm-S} using the state~\eqref{eq-Gibbs}
\begin{eqnarray}
S(T)\approx k_B\left[1-\ln\left(\frac{\hbar\omega(T)}{k_BT} \right) \right],
\label{eq-therm-S-approx}
\end{eqnarray}
is valid in the {\it classical limit} $\hbar\omega(T)/k_BT\ll 1$. The entropy of the oscillator is determined {\rm solely} by the ratio $\hbar\omega(T)/k_BT$, as opposed to the behavior of Var$_X(T)$, Eq.~\eqref{eq-mech-variance-classical}. The temperature dependence of $\omega(T)/T$ dictates the behavior of the entropy, i.e., if $\omega(T)$ grows faster (slower) than linear with $T$ the entropy decreases (increases) with increasing temperature. Due to the monotonic character of the logarithmic function, the local extremes of $\omega(T)/T$ determine the extremes of $S(T)$. The derivative of the approximate result \eqref{eq-therm-S-approx}  
\begin{eqnarray}
\frac{\partial S(T)}{\partial T}=\frac{k_B(2\omega_0^2+a_1T-a_3T^3)}{2T\omega^2(T)},
\label{eq-therm-S-derivative}
\end{eqnarray}
yields a relatively simple sufficient condition for existence of entropy extreme with respect to the temperature $T$.

\begin{figure}[t!]
\centering
\includegraphics[width=.55\linewidth]{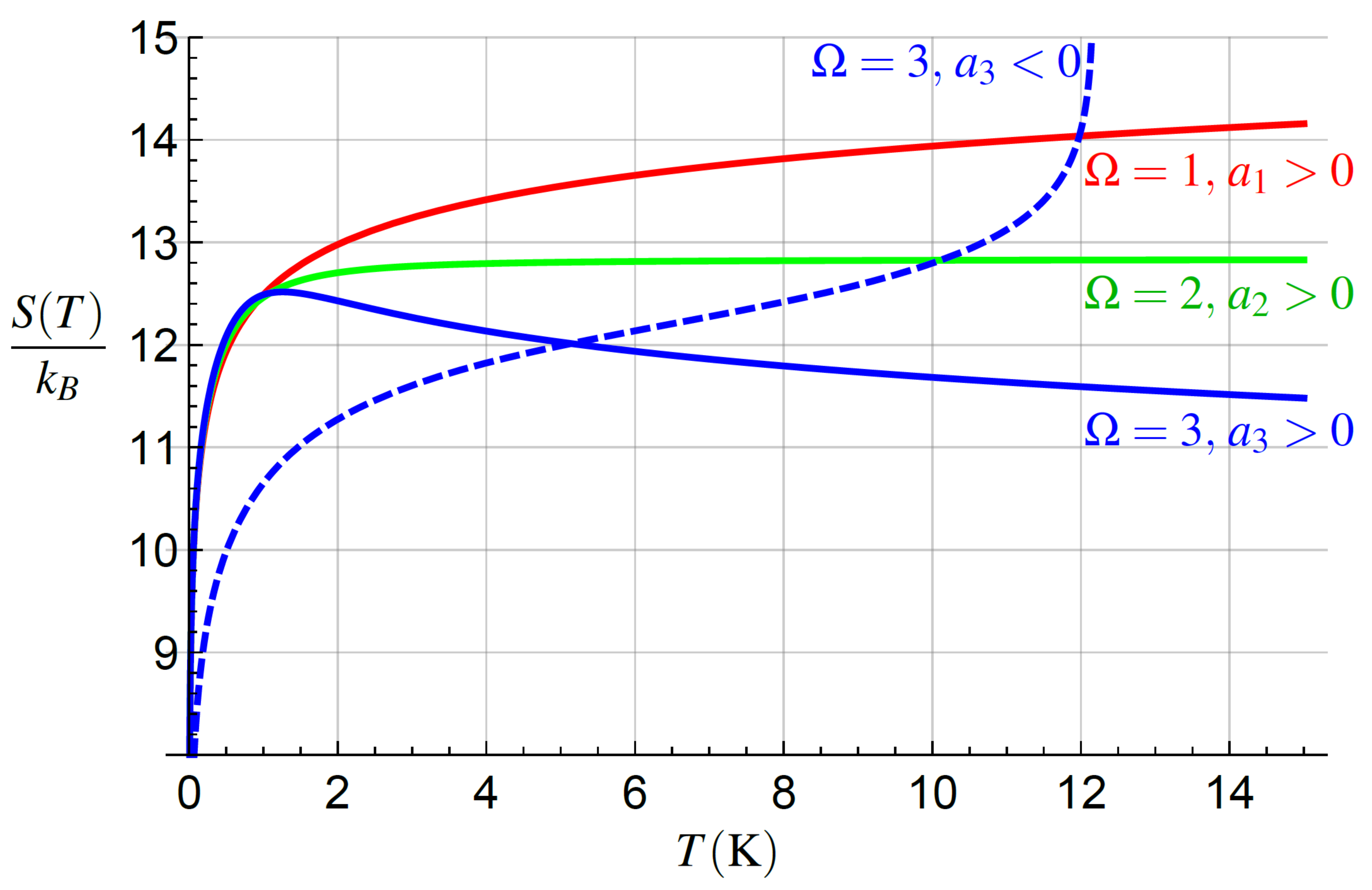}
\caption{The entropy Eq.~\eqref{eq-therm-S} as the function of the temperature for the respective leading-term coefficients $a_\Omega$. For the {\it negative} coefficient $a_3<0$ (blue-dashed) the entropy diverges at points of vanishing $\omega(T)$, $\omega(T_{\rm d})=0$. For {\it positive} leading-term coefficients $a_\Omega$ the behavior depends on the $a_\Omega$ value. In the case $\Omega=1$ (red) the entropy diverges as $S(T)\propto k_B\ln\sqrt{T}$, for $\Omega=2$ (green) the entropy saturates for large $T$ at the value~\eqref{eq-therm-S2-saturation}. For $\Omega=3$ and {\it positive} leading term coefficients $a_3>0$ (blue-full), there exists a local maximum  at the temperature $T_{\rm max}\approx \sqrt[3]{2\omega_0^2/a_3}$ with the maximum value $S(T_{\rm max})$, Eq.~\eqref{eq-therm-S3-max}. In this regime we observe {\em thermal purification} of the oscillator and the entropy  decreases logarithmically. In all plotted cases $|a_\Omega|T \approx\omega_0^2$, $\omega_0\approx 10^6$ rad/s.\cite{aspelmeyer2011} } \label{figure-entropy-T-all}
\end{figure}

For $\Omega=1$,  the entropy $S(T)$ has no local extreme, it monotonically increases, meaning the entropy is unbounded from above, see the dashed line in Fig.~\ref{figure-entropy-T-all}. This is a similar situation as the absence of position variance extreme, discussed in Sec.~\ref{sec:mechanics}. For $a_1>0$ the derivative~\eqref{eq-therm-S-derivative} is positive for all $T$. It is qualitatively similar to the standard case of the thermalized linear oscillator coupled weakly to the heat bath. For the negative leading-term coefficient $a_1<0$ the maximum 
does not exist in the $S(T)$ domain. 

For $\Omega=2$, the entropy has no local extreme, as well. The coefficient $a_2$ does not appear in Eq.~\eqref{eq-therm-S-derivative}, thus in the case $a_2>0$ the entropy $S(T)$ monotonically increases with the temperature $T$ and eventually saturates at the value 
\begin{eqnarray}
\overline{S}\approx k_B\left[1-\ln\left(\frac{\hbar\sqrt{a_2}}{k_B} \right)\right].
\label{eq-therm-S2-saturation}
\end{eqnarray}
Remarkably, this value depends only on the $\omega(T)$ leading term coefficient $a_2$ and not on $\omega_0$. This behavior should be compared to the {\it localization} of the oscillator position $X$, Eq.~\eqref{eq-mech-V2-max}. Contrary to ${\rm Var}_X(T)$ the entropy $S(T)$ saturates, meaning that there is no {\it purification} of the state for the corresponding parameters, although there exists an upper bound on the oscillator entropy, Eq.~\eqref{eq-therm-S2-saturation}.
For a negative leading-term coefficient $a_2<0$ the entropy diverges at the temperature $T_{\rm d}$ at which the frequency vanishes, $\omega(T_{\rm d})=0$. In such case the energy level spacing of the oscillator becomes negligible, causing flat-like population of the levels. 

Finally, if $\Omega=3$ and we assume negative leading-term coefficient $a_3<0$ there exist no $T_{\rm max}>0$ for which the numerator of Eq.~\eqref{eq-therm-S-derivative} vanishes, thus the entropy monotonically increases similarly to the position variance, see Fig.~\ref{fig-mech-all}. Of course, the increase is faster compared to the standard oscillator with a constant frequency. For a positive leading-term coefficient $a_3>0$ the possible points of local extremes are the {\it positive} real roots of the cubic polynomial in Eq.~\eqref{eq-therm-S-derivative}. In the simplified case $a_1\ll a_3T^2$, the extreme appears at  $T_{\rm max}\approx \sqrt[3]{2\omega_0^2/a_3}$, yielding the value according to Eq.~\eqref{eq-therm-S-approx}
\begin{eqnarray}
S(T_{\rm max})\approx k_B\left[1-\ln\left( \frac{\hbar a_3^{1/3}\omega_0^{1/3}}{k_B}\right)\right],
\label{eq-therm-S3-max}
\end{eqnarray}
approximately valid assuming the argument of the logarithm being large enough in accordance with Eq.~\eqref{eq-therm-S-approx}. In this regime, we observe {\it purification} of the oscillator state with increasing temperature, even more unusual phenomena of strong coupling regime. Thus, in this case the thermal localization and purification appears in parallel. It is the best regime to jointly demonstrate both these counter-intuitive phenomena. 
For higher $\omega_0$ the maximum appears at higher temperatures and $S(T_{\rm max})$ is reduced, see Eq.~\eqref{eq-therm-S3-max}.

\section{Heat Capacities}
\label{sec-capacity}

The counter-intuitive phenomena of thermal purification for the von Neumann entropy brought us close to thermodynamic analysis of strong coupling effects. In the classical macroscopic thermodynamics heat capacities yield the amount of heat exchanged between a system and the heat bath when temperature changes during a specific thermodynamic process. For microscopic mechanical oscillator the heat exchange between the oscillator and its environment is hard to measure. In the present work we focus on two capacities (thermodynamic coefficients)
\begin{align}
C_{ S}(T)& =  T \frac{\partial S}{\partial T},\label{eq-thermo-Cs} \\
C_{\mathcal{U}}(T) & = \frac{\partial \mathcal{U} }{\partial T}.\label{eq-thermo-C}
\end{align}
The measurement of the capacities \eqref{eq-thermo-Cs} and \eqref{eq-thermo-C} can provide us information about thermally activated microscopic processes in the system. In the weak coupling limit ($\lambda\to 0$, cf.~Eq.~\eqref{eq-Hmf}), $\omega(T)=\omega_0$ and these quantities are identical. For the oscillator strongly coupled to the bath, each quantity provides a different information and each is a result of different type of possible (at least in principle) measurement. 

First, the entropic capacity, $C_S(T)$, is obtained provided we can reconstruct the equilibrium state $\hat{\rho}(T)$, Eq.~\eqref{eq-Gibbs}, of the oscillator strongly coupled to the bath and calculate the von Neumann entropy. It is experimentally accessible for many optomechanical and electromechanical experiments using the homodyne tomography.\cite{browleyNCom2016} In the case of an oscillator, different approach relying on the ${\rm Var}_X(T)$ measurement is as well possible, being described at the end of this subsection.

Second, the heat capacity $C_{\mathcal{U}}(T)$ is defined through the internal energy function $\mathcal{U}(T)$, given by the difference \cite{TalknerHanggiPRE2016}
\begin{equation}
\mathcal{U}(T) = \langle \op{H}_{\rm tot}\rangle - \langle \op{H}_{\rm B} \rangle_{\rm B},\label{eq-therm-internal-en}
\end{equation}
related to the Hamiltonian of mean force as \cite{TalknerHanggiPRE2016} 
\begin{equation}
\mathcal{U}(T) =  \langle \op{H}(T) \rangle - T \left\langle \frac{\partial \op{H}}{\partial T}  \right\rangle.
\end{equation}

According to Eq.~\eqref{eq-therm-internal-en}, the heat capacity $C_{\mathcal{U}}(T)$ is an outcome of the differential calorimetric measurement. To see this, we note that the right hand side of Eq.~\eqref{eq-therm-internal-en} contains the energy difference between two independent systems:  (i) the oscillator together with the bath, characterized by the total Hamiltonian ${\op{H}_{\rm tot}}$, cf.\ Eq.~\eqref{eq-total-ham}, and (ii) the plain bath without the oscillator with the Hamiltonian ${\op{H}_{\rm B}}$. Because the quantity $\mathcal{U}(T)$ is given by the difference of the average energies of the systems (i) and (ii), its change with temperature, $(\partial \mathcal{U}/\partial T){\rm d}T$, equals to the difference of energy flows into (i) and (ii). Hence, the capacity $C_{\mathcal{U}}(T)$ can be obtained measuring the difference of energy flows into these two systems during temperature changes. It is not necessary to measure the state $\op{\rho}(T)$ directly, however, it requires energy measurement with a rather high precision which is stimulating for further technological development. \cite{PekolaNJP2013,GasparinettiPhysRevApplied2015,GoveniusPRL2016}

The two capacities $C_S(T)$ and $C_{\mathcal{U}}(T)$, can witness the fact that the interaction strength between the oscillator and the bath is beyond the weak-coupling limit. In this limit, both these capacities are strictly positive. The negativity of the capacities reflects strong oscillator-bath coupling. We can formulate the following {\it strong-coupling witness}: If the respective capacity is negative, then, definitely, the system is strongly coupled to its surrounding. Moreover, the capacities can clearly identify the cases when the function $f(T)$, cf.\ Eq.~\eqref{eq-omega}, is a nonlinear function of $T$. 

\begin{figure}[t!]
\centering 
\includegraphics[width=.55\linewidth]{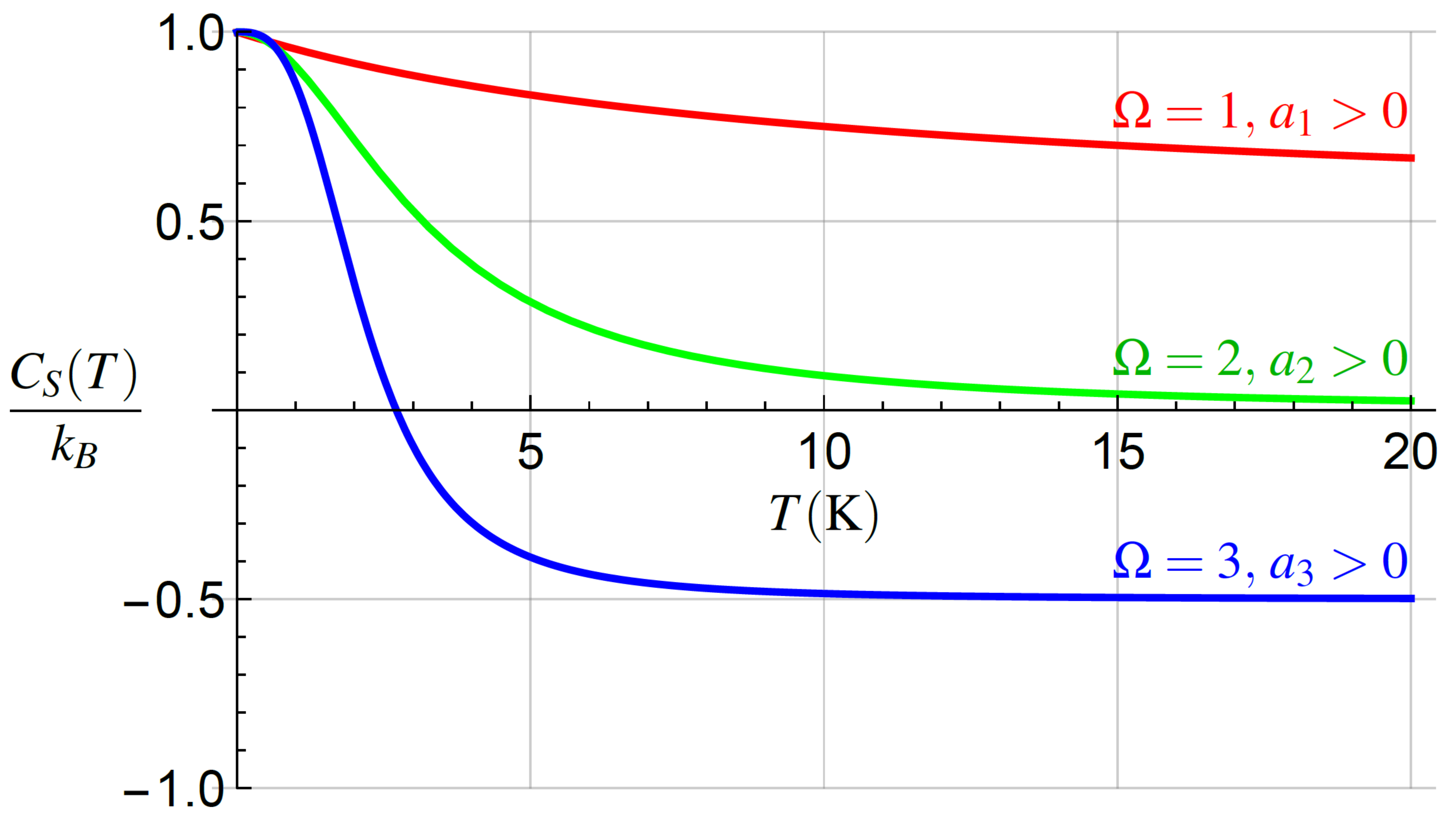}
\caption{The entropic capacity~\eqref{eq-thermo-Cs}, for different values of $\Omega$. For $\Omega\leq 2$, (red, green), $C_S(T)$ approach their asymptotic values $C_S(T)\approx k_B(1-\Omega/2)$ according to Eq.~\eqref{eq-therm-S-derivative}.  For $\Omega\geq 2$, $C_{S}(T)$ may become negative.
In all plotted cases $a_\Omega T\approx\omega_0^2$, $\omega_0\approx 10^6$ rad/s.\cite{aspelmeyer2011} The parameters used are the same as in Fig.~\ref{figure-entropy-T-all}.} \label{figure-capacities-CS-T}
\end{figure}

\begin{figure}[t!]
\centering 
\includegraphics[width=.55\linewidth]{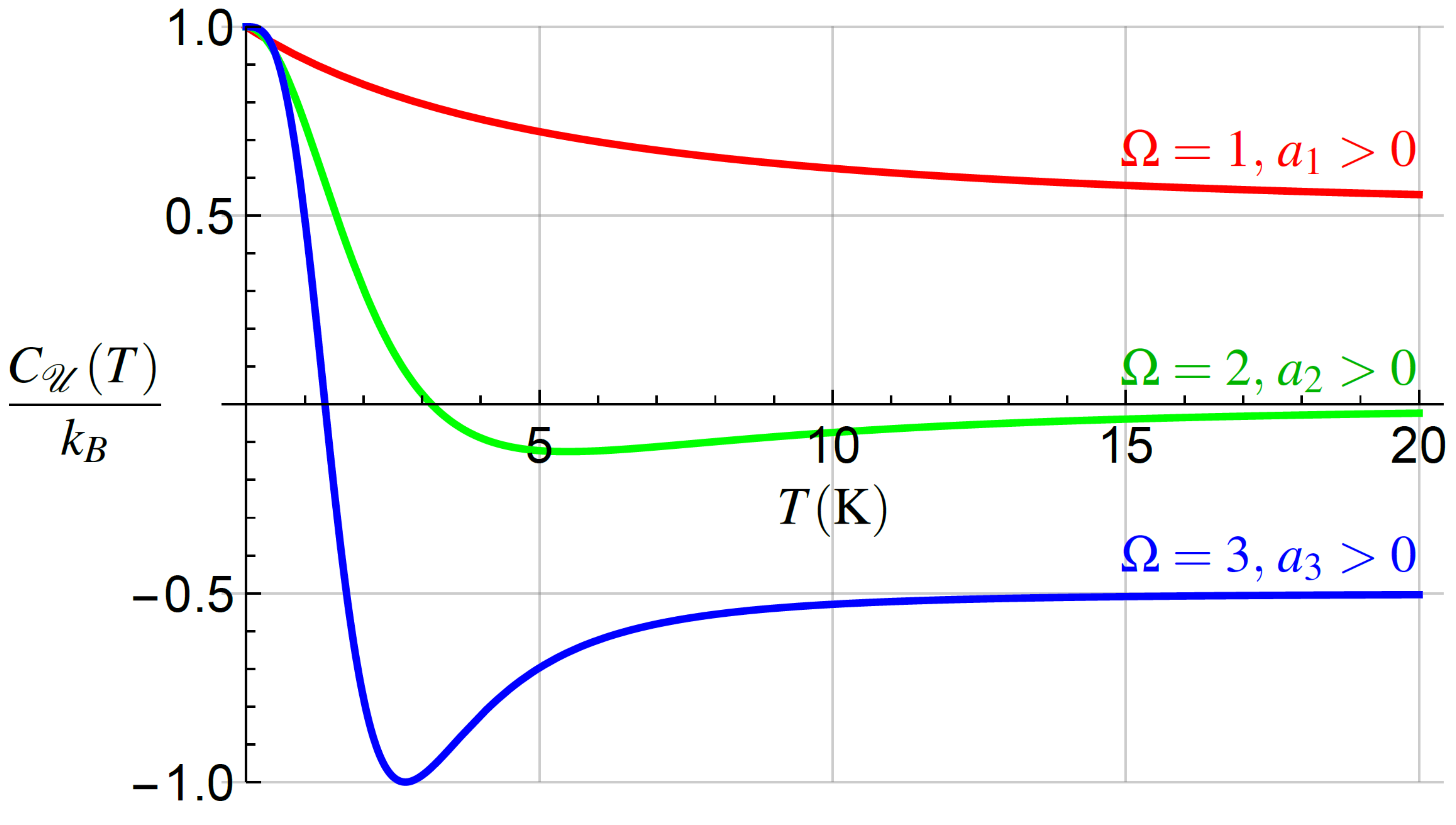}
\caption{The heat capacity~\eqref{eq-thermo-C} for different values of $\Omega$. For $\Omega\geq 2$, $C_{\mathcal{U}}(T)$ becomes negative and in all cases $C_{\mathcal{U}}(T)\leq C_{S}(T)$, cf. Fig.~\ref{figure-capacities-CS-T}. In all plotted cases $a_\Omega T\approx\omega_0^2$, $\omega_0\approx 10^6$ rad/s.\cite{aspelmeyer2011} The parameters used are the same as in Fig.~\ref{figure-capacities-CS-T}.} \label{figure-capacities-C-T}
\end{figure}

\subsection{The Capacity \texorpdfstring{$C_S(T)$}{Cs(T)} }
\label{sub-CS}
The behavior of the entropic capacity $C_S(T)=T\partial S/\partial T$ is rather complex and interesting, cf. Fig.~\ref{figure-capacities-CS-T}, and follows directly from Eq.~\eqref{eq-therm-S-derivative}. For the negative values of the leading-term coefficients $a_\Omega <0$ the capacity diverges at same temperature $T_{\rm d}$ as the entropy. Figure~\ref{figure-capacities-CS-T} shows the capacity for the positive leading-term coefficients, $a_\Omega >0$. For $\Omega\leq 2$, the capacity does not attain negative values as opposed to $\Omega=3$. For larger temperatures, the three plotted curves approach the asymptotic values $C_S\approx k_B(1-\Omega /2)$. The capacity $C_S(T)$ reaches its lowest values for the highest order of the nonlinearity in $T$, $\Omega =3$. For smaller $\omega_0$, while other parameters are kept constant, the value $C_S(T)=0$ is crossed at lower temperatures, allowing for easier observation of $C_S(T)<0$. Negative capacity $C_S(T)$ means that the entropy of our system decreases with the increasing temperature. 

In the classical regime, when both variances are far from the ground-state variance, the equilibrium probability density function factorizes, $\rho(X,P) = \rho(X) \rho(P)$. Then the von Neumann entropy can be decomposed into the position and momentum parts, $S(T)=S_X(T)+S_P(T)$, where $S_X(T)= -k_B\int {\rm d}X\rho(X)\ln[\rho(X)]$, expresses the uncertainty of the probability density $\rho(X)$, and similarly for $S_P(T)$. The entropy $S_X(T)$ can be understood as a counterpart of the position variance ${\rm Var}_X(T)$. Using $S(T)=S_X(T)+S_P(T)$, we obtain $C_S(T)=C_{S_X}(T)+C_{S_P}(T)$, cf. Eq~\eqref{eq-thermo-Cs}. The momentum part contributes by the constant only, $C_{S_P}(T) = k_B/2$. 

In analogy to  the entropic capacity $C_{S_X}(T)$ we can define  the mechanical position-uncertainty coefficient  $C_{{\rm Var}_X}(T)=T (\partial {\rm Var}_X/\partial T)$. From a mechanical viewpoint, it determines a variance which can be reached by heating the oscillator up to a temperature $T$, if that variance constantly increases from $T=0$ with a slope given by its local value $\partial {\rm Var}_X/\partial T$. Its properties are different from $C_{S_X}(T)$. They are related as $C_{S_X}(T)=C_{{\rm Var}_X}(T)/{\rm Var}_X(T)$, thus $C_{{\rm Var}_X}(T)$ can be used to determine $C_S(T)$, without the necessity of estimating the state $\op{\rho}(T)$. In this way, localization of the particle can be directly measured and used to determine the main features of the capacity $C_S(T)$. In turn, this capacity can be used, in our case of an oscillator, to describe the properties of $C_{\mathcal{U}}(T)$, discussed in the next subsection. 

\subsection{The Capacity \texorpdfstring{$C_{\mathcal{U}}(T)$}{Cu(T)} }
\label{sub-CU}
It turns out that the respective capacities $C_S(T)$ and $C_{\mathcal{U}}(T)$ can be related to each other. For instance, we obtain the relation
\begin{equation}
C_{\mathcal{U}}(T)=C_S(T)- T \frac{\partial }{\partial T}\left< \frac{\partial H}{\partial T}\right>,
\end{equation}
yielding another useful result valid
for our HMF~\eqref{eq-oscill-ham}
\begin{eqnarray}\label{eq-thermo-CU-CS-relation}
C_{\mathcal{U}}(T)= \frac{\partial }{\partial T}\left[TC_S(T) \right].
\end{eqnarray}
Clearly, in the weak-coupling limit $C_{\mathcal{U}}(T)$ and $C_S(T)$ are identical. 

Eq.~\eqref{eq-thermo-CU-CS-relation} allows for determination of $C_{\mathcal{U}}(T)$ from $C_S(T)$, achievable by homodyne tomography, \cite{aspelmeyer2011PNAS} or even from the knowledge of ${\rm Var}_X(T)$, see Sec.~\ref{sub-CS}. The behavior of $C_{\mathcal{U}}(T)$ for different values of $\Omega$ is shown in Fig.~\ref{figure-capacities-C-T}, for the same parameter values as used in Fig.~\ref{figure-capacities-CS-T}. 

Fig.~\ref{figure-capacities-C-T} illustrates that $C_{\mathcal{U}}(T)$  can serve for an even more interesting purpose. Surprisingly, it can witness  the order $\Omega$ of the temperature dependence $\omega(T)$ in a more effective way compared to $C_S(T)$, due to the relation $C_{\mathcal{U}}(T)\leq C_S(T)$.
Thus from the behavior of $C_{\mathcal{U}}(T)$  we can determine whether the temperature dependence in \eqref{eq-omega} is linear or of a higher order in $T$. It simultaneously implies, that to reach a strong coupling proved by negative capacity $C_{\mathcal{U}}(T)$, $\omega(T)$ with $\Omega\geq 2$, Eq.~\eqref{eq-omega}, is required. 

Finally, note that we have studied\cite{kolar,kolar2016} the HMF, where the temperature dependence was in the {\it linear} additive term $f(T)\op{X}$, in contrast to the quadratic term $f(T)\op{X}^2$ in the present Eq.~\eqref{eq-oscill-ham}. Such linear term does not lead to $C_{\mathcal{U}}(T)<0$ or $C_S(T)<0$ for any $\Omega\leq 3$, assuming we adopt the same form of $\omega(T)$ as in Eq.~\eqref{eq-omega}. The behavior of the capacity does strongly depend on the particular form of the HMF.

In analogy with classical weak-coupling thermodynamics, one can be tempted to define the third capacity as $C_H(T)=\partial \langle \op{H} \rangle/\partial T$, with $\qmean{\op{H}}$ being the mean value of the HMF, Eq.~\eqref{eq-oscill-ham}.  This quantity is not related to a heat flow between the bath and the oscillator and has no abilities to witness localization and/or purification of the oscillator when the surrounding temperature is changed. However, this quantity resembles the textbook results for the harmonic oscillator heat capacity $C_H(T)=k_{\rm B}$, valid in the weak coupling limit. Only in this limit, $\langle \op{H} \rangle$ has the meaning of the average energy of the system and $C_H(T)$ coincides with the capacities defined in Eqs.~\eqref{eq-thermo-Cs}, \eqref{eq-thermo-C}.

\section{Route to experiment}
\label{sec:conclusions}

We have analyzed mechanical and thermodynamic characteristics of a mechanical oscillator with temperature-dependent energy levels $E_n(T)=\hbar\omega(T)(n+1/2)$. Such model can be interpreted as representing a single light-addressed effective mode of a solid-state part (e.g., membrane or microresonator) of an optomechanical device. Such-solid state structures have typically their basic characteristics (like frequency) dependent on the temperature of its surroundings, $\omega=\omega(T)$.  We used this link with our model and together with interpretation of such temperature-dependence as a consequence of strong coupling of the mode to the ambient heat bath, we have discussed the aforementioned mechanical and thermodynamic properties stemming from such dependence. For different functional forms of this dependence we show that in a high-temperature regime, we encounter a wide variety of qualitatively different behaviors regarding the oscillator localization (reduction of its position variance), Sec.~\ref{sec:mechanics}, the purification (reduction of its von Neumann entropy), Sec.~\ref{sec-entropy}, or ability to witness strong oscillator-bath coupling, Sec.~\ref{sec-capacity}. The main result of our paper is that for the {\it linear} temperature dependence of the square of the oscillator frequency we can not observe any of these effects. On contrary, for the {\it cubic} temperature dependence of the squared frequency, both effects appear simultaneously and are more pronounced with increasing temperature, cf. Figs.~\ref{fig-mech-all}-\ref{figure-entropy-T-all}. 

Regarding the frequency ranges suitable for the observation of interesting effects described in this paper, we conclude the following. When localization, state purification, and/or witnessing strong coupling to the bath are in the focus of interest, it is preferable to use oscillators with lower bare frequencies. On the other hand, if the figure of merit is to keep the oscillator in the low-entropy regime, one should preferably work with a higher bare frequency oscillators.

As the first step towards experimental observation of the predicted strong coupling effects can be design of a simulator where both temperature and frequency are tuned in a correlated way. This simulator can be practically realized in a similar way as the classical squeezing of a thermal state of levitating nanoparticle out of the equilibrium was done,\cite{RashidPRL117} if the frequency change would be correlated with heating of the nanoparticle. To test the quantum non-equilibrium version of this effect,  mechanical state of trapped ion with uncertainty squeezed below the ground state can be used.\cite{2018arXiv181201812B} Alternatively, a transient time-dependent squeezing of light can be considered.\cite{marek2014} Such simulator will allow not only to verify our current results, but also to simulate further topics of stochastic and quantum thermodynamics, such as the performance of quantum heat engines at strong coupling,\cite{EisertNJP2014, Klimovsky2015, Kosloff2016, NewmanPRE2017, LlobetPRL2018} including impact of fluctuations on their efficiencies,~\cite{HolubecRyabovPRL2018} and the effects of thermally induced coherence.\cite{GuarnieriPRL2018} The result of these simulations will explore the possibility to measure such effect in different experimental regimes. Next step is technological, looking for a design of mechanical oscillator with maximized temperature-dependent frequency at a thermal equilibrium and minimal noise and anharmonicity.

\bibliography{references_resubmission}


\section*{Acknowledgements}

M.K.\ and R.F.\ acknowledge the support of the project GB14-36681G of the Czech Science Foundation. A.R.\ gratefully acknowledges support of the project No.~17-06716S of the Czech Science Foundation. R.F. have received national funding from the MEYS of the Czech Republic (Project No. 8C18003) under Grant Agreement No. 731473 within the QUANTERA ERANET cofund in quantum technologies implemented within the European Unions Horizon 2020 Programme (Project TheBlinQC).  A.R.\ and M.K.\ are grateful to P.\ Cabi{\v s}ov{\' a} for stimulating discussions. 

\section*{Author contributions statement}

All authors developed the concept, performed the theoretical analysis, and contributed substantially to the content of this manuscript. All authors reviewed the manuscript. 

\section*{Competing interests}

The authors declare no competing interests.

\end{document}